\documentclass{article}
\usepackage{graphicx} 

\usepackage{amsmath,amssymb}
\usepackage{hyperref}
\usepackage{subcaption}
\usepackage{bbm}
\usepackage{adjustbox}
\usepackage[numbers,compress]{natbib}

\input{preamble}

\newcommand{\der}[0]{\mathrm{d}}

\newcommand{\figsize}{0.7}

\title{Computing Nonequilibrium Responses with Score-shifted Stochastic Differential Equations}

\author{J\'{e}r\'{e}mie Klinger and Grant M. Rotskoff}
\date{\today}

\begin{document}

\maketitle

\begin{abstract}
    Using equilibrium fluctuations to understand the response of a physical system to an externally imposed perturbation is the basis for linear response theory, which is widely used to interpret experiments and shed light on microscopic dynamics. 
    For nonequilibrium systems, perturbations cannot be interpreted simply by monitoring fluctuations in a conjugate observable---additional dynamical information is needed. 
    The theory of linear response around nonequilibrium steady states relies on path ensemble averaging, which makes this theory inapplicable to perturbations that affect the diffusion constant or temperature in a stochastic system. 
    Here, we show that a separate, ``effective'' physical process can be used to describe the perturbed dynamics and that this dynamics in turn allows us to accurately calculate the response to a change in the diffusion.
    Interestingly, the effective dynamics contains an additional drift that is proportional to the ``score'' of the instantaneous probability density of the system---this object has also been studied extensively in recent years in the context of denoising diffusion models in the machine learning literature.
    Exploiting recently developed algorithms for learning the score, we show that we can carry out nonequilibrium response calculations on systems for which the exact score cannot be obtained. 
\end{abstract}

\paragraph{Introduction}

Using the fluctuation dissipation relation to compute the response of a physical system at equilibrium to an externally applied force relies on Onsager reciprocity~\cite{onsager_reciprocal_1931,kubo_fluctuation-dissipation_1966}.
The assumptions of reciprocity are violated when a system is driven away from equilibrium, leading to response theories that require more explicit information about the underlying microscopic dynamics~\cite{maes_fluctuation-response_2013}.
The general formalism for computing responses around nonequilibrium stationary states (NESSs) relies on relating the perturbed path ensemble to the stationary path ensemble, which only in certain cases can be directly formulated as a fluctuation dissipation relation~\cite{chun_nonequilibrium_2021}.
This framework is, however, mathematically delicate; the formal approach breaks down when we perturb the temperature or the diffusivity of the driving stochastic process. 
This article describes a straightforward transformation that alleviates the difficulties that arise in such cases. 

Despite the importance of understanding response to changes in the diffusivity for stochastic systems in active matter~\cite{bechinger_active_2016, chennakesavalu_probing_2021}, biophysics~\cite{brown_theory_2020, yamamoto_dynamic_2017}, and nanoscale devices~\cite{brandner_thermodynamics_2015, saha_information_2023}, it remains difficult to compute the sensitivity---the differential response of an observable to a perturbation---when we alter either the temperature or diffusivity in a stochastic, nonequilibrium dynamics.
A number of works have sought to clarify these difficulties: \citet{Falasco2016} show that, time-dependent (but not space-dependent) perturbations in the diffusion tensor correspond to a reparameterization of time; \citet{brandner_thermodynamics_2015} considers the case of periodic perturbations in temperature; \citet{baiesi_thermal_2014} introduces a high-frequency regularization in the noise to compare stochastic trajectories. 
Alongside these strategies in the physics literature, Malliavin calculus~\cite{Nualart} provides a precise mathematical structure to compute arbitrary sensitivities.
\citet{warren_malliavin_2012} used this approach for sensitivities to perturbations in the drift, where the approach coincides with the usual Girsanov transformation. 
While computing sensitivities to space and time-dependent diffusion tensors via the Malliavin calculus has been used in financial mathematics, implementing Malliavin derivatives for physical dynamics requires a complicated accounting of several inter-related stochastic processes, leading to practical constraints on its applicability~\cite{Fournie1999}. 

Here, we show that a transformation of a diffusion-perturbed SDE resuscitates the path reweighting approach. Specifically, we define an effective stochastic process with the \textit{unperturbed diffusion tensor}, whose governing Fokker Planck equation evolves the initial system in the same way as the diffusion-perturbed process.
This alternative perturbed equation involves the ``score'', which is the gradient of the log of the instantaneous distribution of the system. While for most nonequilibrium systems the ``score'' cannot be computed analytically, recent progress in the machine learning literature on score-based diffusion models~\cite{song_score-based_2022} and score estimation~\cite{Boffi2023, hyvarinen_estimation_2005} have led to new computational tools that provide arbitrarily precise score estimates, which would have previously seemed unobtainable. 
As we show analytically and numerically below, this transformed SDE gives excellent results for diffusion sensitivity on a variety of systems, including equilibrium, stationary nonequilibrium, and non-stationary nonequilibrium cases.

\paragraph{Computing diffusion sensitivies with the score-shifted SDE.---}
We consider a system with coordinates $\xb \in \Omega \subset \RR^d$ and represent the dynamics as an overdamped stochastic differential equation
\begin{equation}
	\der \Xb_t = \bb(\Xb_t, t )\der t + \sigma(\Xb_t, t) \der \Wb_{t},
    \label{eq:osde}
\end{equation}
where $\bb:\RR^d \times [0, T] \to\RR^d$ is the drift, $\sigma: \RR^d \times [0, T] \to \RR^{d\times m}$ is a space and time-dependent mobility tensor, and $\Wb$ is an $m$-dimensional Wiener process.
We want to compute the \emph{sensitivity} of an arbitrary, bounded, continuous observable $A\in \mathcal{C}_{\rm b}^0(\Omega,\RR)$ to a perturbation to $\sigma$, 
\begin{equation}
    \partial_{\tilde{\sigma}}A_t = \lim_{\epsilon\to 0} \epsilon^{-1} \left[\avg{A(t)}_{\epsilon} - \avg{A(t)}\right]
    \label{eq:sens}
\end{equation}
where $\avg{\cdot}$ denotes an expectation with respect to \eqref{eq:osde} and $\avg{\cdot}_{\epsilon}$ denotes an average with respect to the perturbed SDE, 
\begin{equation}
    \der \Xb^{\epsilon}_t = \bb(\Xb^{\epsilon}_t, t )\der t + \left[ \sigma(\Xb^{\epsilon}_t, t)  + \epsilon \tilde{\sigma}(\Xb^{\epsilon}_t, t) \right]  \der \Wb_{t}.
    \label{eq:psde}
\end{equation}
The primary challenge when trying to compute a sensitivity of the type defined in \eqref{eq:sens} is that widely-used path reweighting techniques fail~\cite{maes_fluctuation-response_2013, warren_malliavin_2012, Baiesi2013, chun_nonequilibrium_2021}. 
Because the path measures of the processes defined by $\Xb_t$ and $\Xb_t^{\epsilon}$ are mutually singular, a straightforward application of the Girsanov theorem cannot be carried out~\cite{oksendal_stochastic_2003}. 
There is, however, a simple and surprisingly convenient remedy.  

To build a stochastic process that can be reweighted relative to~\eqref{eq:osde}, we let $\Sigmasf(\xb,t) = \sigma(\xb,t) \Tilde{\sigma}^T(\xb,t) + \sigma^T(\xb,t) \Tilde{\sigma}(\xb,t)$ and consider instead the SDE,
\begin{equation}
    \der \Xb^{s}_t = \left[\bb(\Xb^{s}_t, t ) - \frac{\epsilon}{2}\bigl( (\Sigmasf(\Xb^s_t,t) s(\Xb^s_t,t) + \nabla : \Sigmasf(\Xb^s_t,t)\bigr)\right]\der t +  \sigma(\Xb^{s}_t, t) \der \Wb_{t},
    \label{eq:psde2}
\end{equation}
and we use the notation
\begin{math}
    \left[\nabla : \mathsf{G} \right]_i = \sum_{j} \frac{\partial}{\partial \xb_j} \mathsf{G}_{ij}
    \label{eq:contract}
\end{math}
and 
\begin{math}
    \nabla \nabla : \mathsf{G}  = \sum_{ij} \frac{\partial^2}{\partial \xb_i \partial \xb_j} \mathsf{G}_{ij}
    \label{eq:diad}
\end{math}
to denote the double dot products of a differential operator with a dyadic tensor~\cite{pavliotis_stochastic_2014}. 
The solution of the associated Fokker-Planck equation,
\begin{equation}
\label{eq:pfke2}
\partial_t \rho^s_t = -\nabla \cdot [ (\bb(\cdot, t ) \rho_t^s ] + \frac12 \nabla \nabla : \left[ (\mathsf{D} + \epsilon \Sigmasf) \rho_t^s \right]
\end{equation}
with 
\begin{math}
\label{eq:diff_tensor}
\mathsf{D}(\xb,t) = \sigma(\xb,t)\sigma^T(\xb,t)
\end{math}
defines $s$, the ``score'' function that appears in \eqref{eq:psde2},
\begin{equation}
    \label{eq:instantaneous_score}
    s(\xb,t) = \nabla_{\xb} \log \rho^s_t(\xb).
\end{equation}

In App.~\ref{app:weak}, we show that $\rho_t^s$ has controlled error to first order in $\epsilon \to 0$ compared with the solution $\rho_t^{\epsilon}$ of the Fokker-Planck equation of the original, perturbed dynamics \eqref{eq:psde}, which solves
\begin{equation}
\label{eq:pfke3}
\partial_t \rho^\epsilon_t = -\nabla \cdot [ (\bb(\cdot, t ) \rho_t^\epsilon ] + \frac12 \nabla \nabla : \left[ \tilde{\mathsf{D}} \rho_t^\epsilon \right],
\end{equation}
with \begin{math}
\label{eq:pdiff_tensor}
\tilde{\mathsf{D}}(\xb,t) = (\sigma(\xb,t) + \epsilon \tilde{\sigma}(\xb,t))(\sigma(\xb,t) + \epsilon \tilde{\sigma}(\xb,t))^T.
\end{math}
In particular, this allows us to compute the sensitivity of an observable $A$ with respect to a perturbation using expectations over the modified process~\eqref{eq:psde2}, i.e., 

\begin{equation}
    \partial_{\tilde{\sigma}}A_t = \lim_{\epsilon\to 0} \epsilon^{-1} \left[\avg{A(t)}_s - \avg{A(t)}\right].
    \label{eq:sens2}
\end{equation}
Conveniently, this new process is constructed so that path reweighting relative to the original dynamics~\eqref{eq:osde} using Girsanov can be carried out.
A straightforward expansion to first order in $\epsilon$ yields,
\begin{equation}
\label{eq:sens3}
\begin{split}
    \avg{A(t)}_s &= \avg{A(t) \frac{\dP^s}{\dP}}\\
    &= \avg{A(t) \exp\left[- \epsilon \int_0^T \sigma(\Xb_u,u)^{-1} F(\Xb_u,u)\der \Wb_u + \frac{\epsilon ^2}{2} \int_0^T ||\sigma(\Xb_u,u)^{-1} F(\Xb_u,u)||^2\der u \right]}\\
    &\underset{\epsilon \to 0}{=}\avg{A(t)}- \epsilon \avg{A(t) \int_0^T \sigma(\Xb_u,u)^{-1} F(\Xb_u,u)\der \Wb_u } + \mathcal{O}(\epsilon^2), \\
\end{split}
\end{equation}
and
\begin{equation}
\label{eq:pforce}
    F(\xb,t) \underset{\epsilon \to 0}{=}  \frac{1}{2}\left(\Sigmasf(\xb,t) s^0(\xb,t) + \nabla : \Sigmasf(\xb,t)\right) +\mathcal{O}(\epsilon^2),
\end{equation}
where $s^0$ is the score of the instantaneous distribution of~\eqref{eq:osde}. 
Our main result is a simple expression for the sensitivity to an arbitrary perturbation in the diffusivity,
\begin{equation}
\label{eq:sensfinal}
    \partial_{\Tilde{\sigma}}A_t = -\frac{1}{2}\avg{A(t) \int_0^t \sigma(\Xb_u,u)^{-1}\left[\Sigmasf(\Xb_u,u) s^0(\Xb_u,u) + \nabla : \Sigmasf(\Xb_u,u)\right]\der \Wb_u },
\end{equation}
which can be computed along the dynamics of the unperturbed process $\Xb_t$. We emphasize that the formula~\eqref{eq:sensfinal} is extremely general, valid for both equilibrium and nonequilibrium dynamics, and making no assumptions of stationarity. 
For both overdamped and underdamped equilibrium dynamics, we show that making use of the explicit score in~\eqref{eq:sensfinal} recovers the fluctuation-dissipation theorem (FDT). However, for general non equilibrium systems, we must approximate the unknown score. We leverage recent breakthroughs in machine learning that have led to robust strategies for estimating the score from data~\cite{song_score-based_2022, Boffi2023,hyvarinen_estimation_2005} and provide a route to computing sensitivities for nonequilibrium systems. 
In what follows, we demonstrate the numerical and analytical validity of the estimator in a variety of examples, and highlight excellent performance in comparison with FDTs, finite differencing and explicit Malliavin sensitivities.

\paragraph{Numerical examples.---}
We begin by using~\eqref{eq:sensfinal} to evaluate the sensitivity of an equilibrium Gibbs-Boltzmann distribution to a perturbation in the temperature $T$, i.e., 
\begin{equation}
\label{eq:odeq}
    \begin{split}
        \der \Xb_t = - \nabla U(\Xb_t)\der t + \sqrt{2 (T+\delta T)} \der \Wb_t
    \end{split}
\end{equation}
with $X_0 \sim \rho_{\rm eq}(\xb) \propto \exp\left[-U(\xb)/T\right]$. \footnote{To first order in $\delta T \equiv \epsilon$, this is strictly equivalent to evaluating the sensitivity $\partial_{\Tilde{\sigma}}A_t $ of the perturbed SDE
\begin{math}
\label{eq:od1}
        \der \Xb_t = - \nabla U(\Xb_t)\der t + (\sigma + \epsilon \tilde{\sigma})\der \Wb_t
\end{math}
where $\sigma(\xb, t) = \sqrt{2 T}$ and $\tilde{\sigma}(\xb, t) = 1/(\sqrt{2 T})$ are simple scalars.} 
Making use of the general sensitivity formula \eqref{eq:sensfinal} and the explicit unperturbed score $s^0(\xb,t) = -\nabla U(\xb)/T$ we obtain the equilibrium sensitivity
\begin{equation}
\label{eq:od3}
\partial_{\Tilde{\sigma}} A_t =  \frac{1}{\sqrt{2 T^3}}\avg{A(t) \int_0^t \nabla U(\Xb_s)\der \Wb_s}.
\end{equation}
We show in App.~\ref{app:eq-od} that the overdamped equilibrium sensitivity \eqref{eq:od3} is equivalent to the standard fluctuation dissipation relation \cite{kubo_statistical_1991, Baiesi2013}
\begin{equation}
\label{eq:fdt1}
\frac{1}{\sqrt{2 T^3}}\avg{A(t) \int_0^t \nabla U(\Xb_s)\der \Wb_s}=\frac{1}{T^2}\left[\avg{A U}-\avg{A(t)U(0)}\right]. 
\end{equation}
Extending the equilibrium sensitivity \eqref{eq:od3} to perturbed underdamped equilibrium dynamics with friction coefficient $\gamma$ is straightforward and the temperature sensitivity is given in App.~\ref{app:eq-od}.

\begin{figure}[h!]
    \centering
    \includegraphics[width = \figsize\textwidth]{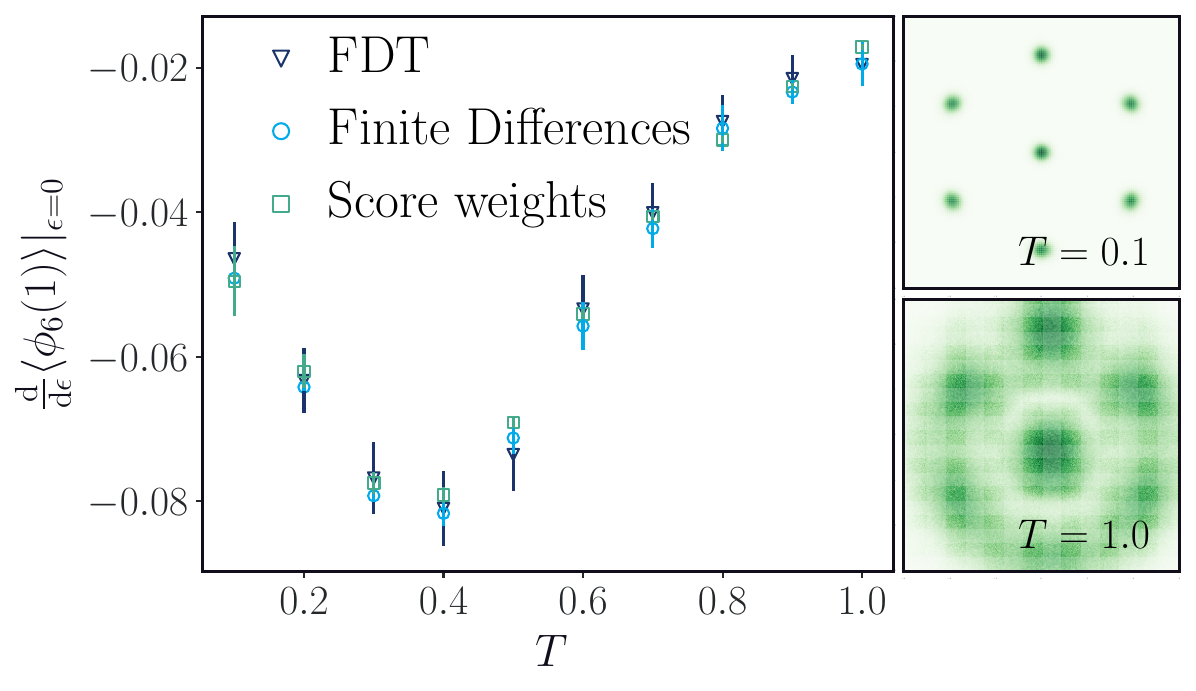}
    \caption{Temperature response for 7 Lennard-Jones interacting particles ($\sigma = 2.0$, $\epsilon = 1.0$) in a harmonic confining trap. We compute the sensitivity of the $\phi_6$ order parameter at $t=1$ after turning the perturbation on. The FDT, finite difference and score sensitivity estimates are respectively averaged over $10^6$, $4.10^6$ and $2.10^7$ samples. Right panel : aligned histograms for high and low temperatures. The hexatic nature of order is most present at low $T$, while diffusion prevails at high $T$.}
    \label{fig:lj}
\end{figure}

As an illustration, we consider an underdamped equilibrium system of 7 interacting Lennard-Jones particles with pairwise interaction potential 
\begin{math}
U_{\rm LJ}(r) = 4 \epsilon \left[\left(\frac{\sigma}{r}\right)^{12}-\left(\frac{\sigma}{r}\right)^{6}\right]
\label{eq:lj1}
\end{math}
confined in a simple harmonic potential
\begin{math}
U_{\rm el}(\xb) = (\xb\cdot\xb)/2.
\label{eq:lj2}
\end{math}
We display on Fig. \ref{fig:lj} the finite-time sensitivity of the particle-averaged hexatic order parameter 

\begin{equation}
    \phi_6(\xb) = \frac{1}{7} \sum_{i=1}^7 \left|\sum_{j \neq i} e^{6 \boldsymbol{i} \theta_{ij}}\right|
\end{equation}
and compare the underdamped score based sensitivity \eqref{eq:app-ud2} to both finite difference and fluctuation dissipation estimates of the thermal sensitivity of $\phi_6$ at a fixed time, $\partial_T \phi_6(t=1)$. Note that we observe good agreement over a range of temperatures spanning both typical ordered (low $T$) and disordered (high $T$) states.

\paragraph{Perturbing around a nonequilibrium steady-state.---}
For a general nonequilibrium steady-state (NESS) system, the FDT breaks down and a lack of explicit form for the stationary distribution of the system prevents direct evaluation of static sensitivities. 
Recent progress on score-estimation using flexible neural networks~\cite{song_score-based_2022,Boffi2023,hyvarinen_estimation_2005} has enabled computation of $\nabla\log \rho_{\rm ss}$ for nonequilibrium distributions that were previously intractable. 
Leveraging these algorithms allows us to build a precise representation of the NESS score which, together with the score sensitivity formula \eqref{eq:sensfinal} constitute unbiased estimators of diffusion sensitivities. 

\begin{figure}[h!]
    \centering
    \includegraphics[width = \figsize\textwidth]{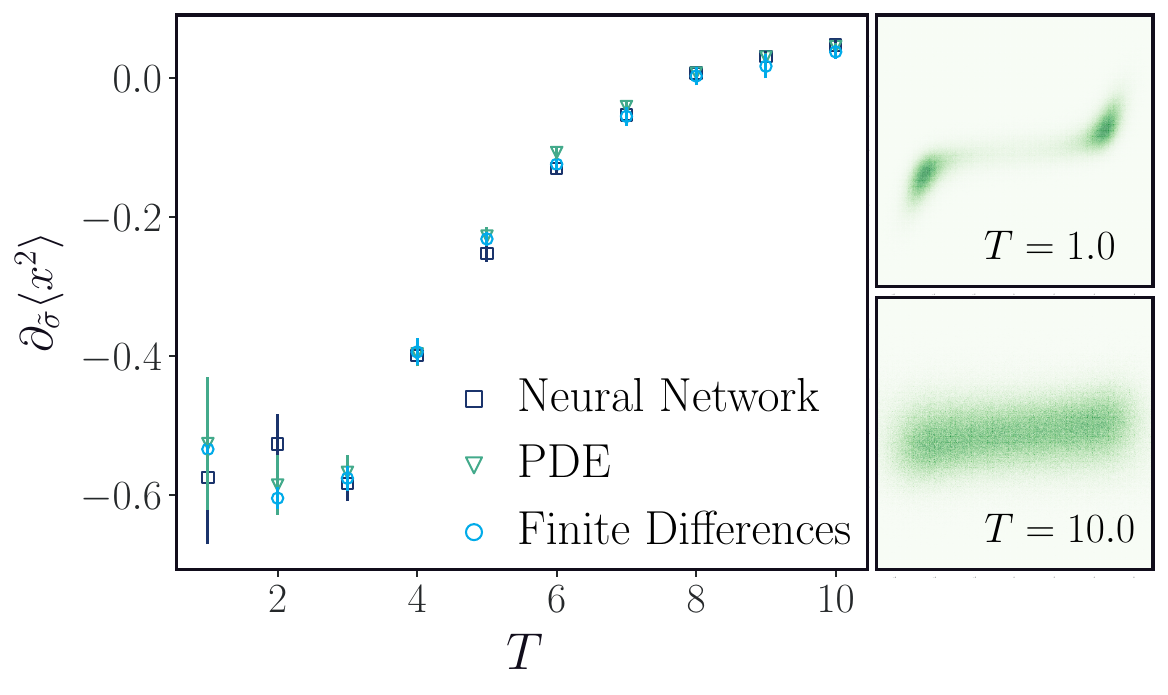}
    \caption{Static temperature response of the stationary variance associated to the distance between two interacting active particles on a ring. The finite difference,  neural network and numerical solution estimates of the variance sensitivity are respectively averaged over $10^6$, $10^6$ and $4.10^6$ samples. 
    Right panel : histograms in the $(x,g)$ plane for high and low temperatures. Probability localization is most present at low $T$, while diffusion prevails at high $T$.
    }
    \label{fig:ness}
\end{figure}

As an illustration, we consider in the following a four dimensional NESS system initially introduced in \cite{Boffi2023} and closely follow their notations for clarity. The system is composed of two one-dimensional interacting active particles on a ring of size $L$ whose positions $x^i$ and orientations $g^i$ evolve according to the coupled SDEs,
\begin{equation}
\begin{split}
\label{eq:ness1}
    &\der x^i_t = \left[\mu f(x^i_t-x^j_t) + v_0 g^i_t\right]\der t +\sqrt{2 T}\der W^{i,x}_t\\
    &\der g^i_t = -\gamma g^i_t \der t +\sqrt{2 \gamma}\der W^{i,g}_t
\end{split}
\end{equation}
where $W^{i,k}_t$ and $W^{j,l}_s$ are independent Wiener processes. The soft repulsive interaction kernel is given by
\begin{equation}
\label{eq:ness2}
    f(x)= \frac{x}{\alpha |x|}\log\left(1+\exp(\alpha(2r - |x| )\right),
\end{equation}
where $\alpha$ and $r$ are parameters accounting respectively for the intensity and cutoff distance of the interaction. 
Because of periodicity, the system is fully described by the reduced variables $x_t = x^1_t -x^2_t $, $g_t = g^1_t - g^2_t $ and the corresponding dynamics
\begin{equation}
\begin{split}
\label{eq:ness3}
    &\der x_t = \left[2 \mu f(x_t) + v_0 g_t\right]\der t + 2 \sqrt{T}\der W^x_t\\
    &\der g_t = -\gamma g_t \der t + 2 \sqrt{ \gamma}\der W^g_t,
\end{split}
\end{equation}
where $W^{i}_t$ and $W^{j}_s$ are independent Wiener processes.
As discussed in depth in \cite{Boffi2023}, the variable $(x_t,g_t)$ reaches a NESS (see Fig. \ref{fig:ness}), whose unknown stationary distribution $\rho_{\rm ss}$ depends on the temperature $T$. 

To make use of the sensitivity formula~\eqref{eq:sensfinal}, we represent the stationary score $s^0$ with a neural network trained on stationary data (see App.~\ref{app:active-training} for training details), and we numerically evaluate the static sensitivity of the variance $\avg{x^2}$ of the reduced system \eqref{eq:ness3} with respect to a change in temperature $T$, as shown in Fig.~\ref{fig:ness}.
The measured sensitivity agrees within error to both an estimate via finite differences and also by numerical solution of the associated Fokker-Planck equation.
This agreement holds across a range of temperatures, but higher estimation errors at lower $T$ arise due to stronger gradients of $\rho_{\rm ss}$ (see right panel of Fig~\ref{fig:ness}), which are computationally more difficult to represent.

\paragraph{Nonequilibrium non-stationary dynamics.---}
While stationarity is a typical assumption for physical perturbation theories, there are many settings that occasion computing sensitivities for non-stationary dynamics.
Specifically, throughout the financial risk literature, diffusion sensitivities are studied for their implications in options pricing, often using the Malliavin calculus. 
Our framework applies directly to the non-stationary setting, though we expect that learning a non-stationary score function for an arbitrary dynamics will be computationally challenging. 
As a proof of concept, we apply the score sensitivity formula \eqref{eq:sensfinal} to evaluate diffusion sensitivities associated with variations of the volatility $\sigma$, or \textit{vega}  \cite{Fournie1999}, in the celebrated one-dimensional Black-Scholes \cite{bs-73} model, which is defined by the stochastic differential equation
\begin{equation}
\label{eq:bs1}
		\der X_t = \mu X_t \der t + \sigma X_t \der W_t;\hspace{10pt}X_0 = 1.
\end{equation}

\begin{figure}[h!]
    \centering
    \includegraphics[width = \figsize\textwidth]{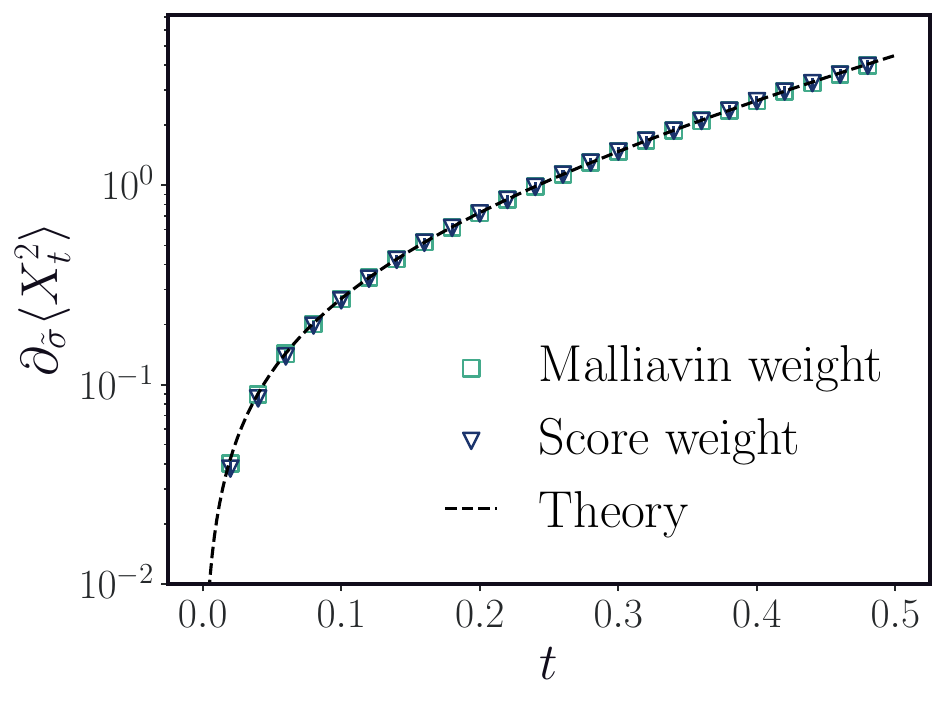}
    \caption{Black-Scholes model with $\mu=1, \sigma=1$; time dependent \textit{vega} of the second moment $\partial_{\tilde{\sigma}}\avg{X_t^2}$.  Both Malliavin and score estimates are averaged over $4.10^5$ sample trajectories. The dashed analytical line is given in App.~\ref{app:malliavin}.}
    \label{fig:bs}
\end{figure}

Because the solution to \eqref{eq:bs1} is an exponential Brownian Motion
\begin{math}
 \label{bs2}
		X_t = \exp\left[(\mu-\frac{\sigma^2}{2})t + \sigma W_t\right],
	\end{math}
the unperturbed dynamical score is explicitly given by
\begin{math}
\label{eq:bs-score}
	s^0(x,t) = -\frac{3}{2 x}+\frac{\mu}{x \sigma^2	}-\frac{\log(x)}{\sigma^2 t x}.
\end{math}
In the score sensitivity formula~\eqref{eq:sensfinal}, the volatility sensitivity \textit{vega} corresponds to $\partial_{\Tilde{\sigma}}A_t$ with
\begin{math}
\label{eq:bs-sigmat}
		\Tilde{\sigma}(x) = x
\end{math}
and we immediately obtain
\begin{equation}
 \label{eq:bs-sens}
		\partial_{\Tilde{\sigma}}A_t =  \avg{A(t)\left[\int_{0}^{t} \frac{W_s}{\sigma s}\der W_s- W_t\right]}.
\end{equation}
We prove in App.~\ref{app:malliavin} that the score sensitivity~\eqref{eq:bs-sens} is equal to the Malliavin \textit{vega} \cite{Fournie1999} 
\begin{equation}
\label{eq:bs-mav}
\avg{A(t)\left[\frac{W_t^2}{\sigma t} - W_t -\frac{1}{\sigma}\right]},
\end{equation}
and numerically compare both estimates of $\partial_{\tilde{\sigma}}\avg{X_t^2}$ on Fig. \ref{fig:bs}. Our estimate \eqref{eq:bs-sens} is in perfect agreement with both the Malliavin result and the explicit analytical sensitivity.

\paragraph{Conclusions.---} Predicting the response of a nonequilibrium physical system to an externally imposed change can elucidate subtle, dynamical fluctuations that underlie the sensitive and surprising adaptability of nanoscale systems. 
The challenge of carrying out these calculations in physical systems of interest, however, has led to a variety of different approaches that are not applicable in every setting. 
The general methodology that we have introduced enables us to leverage new computational tools from machine learning to compute the response to a perturbation in the diffusion tensor of a stochastic dynamics.
For equilibrium systems, the score that appears in~\eqref{eq:sensfinal} is given exactly in terms of the potential energy and we recover the classical fluctuation dissipation theorem.
For nonequilibrium systems, both at stationarity and otherwise, the score can be estimated and represented with a neural network.
Approximating the score in this way gives excellent results on a variety of systems and offer considerable flexibility when compared with the Malliavin approach. 

Examining the score-perturbed SDE~\eqref{eq:psde2} leads to a clear physical picture of the effective process: the additional drift in this equation acts as a force to enhance or suppress the variance of the instantaneous distribution, depending on the nature of the perturbation. 
While we have shown that this strategy exactly coincides with the Malliavin approach for the important case of Black-Scholes, a more general proof of equivalence remains.
Because the framework that we have introduced simplifies computations of gradients with respect to the diffusivity for an arbitrary SDE, it also enables optimization, which we anticipate will be useful in the general machine learning context. 

\appendix

\section{Controlling the error between the score-shifted and the perturbed processes} \label{app:weak}

In this appendix, we formally analyze the discrepancy between the score-shifted process and the exactly perturbed process. 
We make assumptions that lead to well-posed partial differential equations, for example, we assume the assumption of a uniformly parabolic diffusion tensor. 
\begin{assumption} \label{assump:linfty}
    The drift and mobility functions $\bb_i, \sigma_{ij}, \tilde{\sigma}_{ij}$ are elements of the function space $L^{\infty}(\Omega)$. 
\end{assumption}
\begin{assumption} \label{assump:parab}
For all $\epsilon > 0$ and $\xib\in \RR^d$ the mobility function $\sigma:\Omega \to \RR^d$ and the perturbation $\tilde\sigma: \Omega \to \RR^d$ satisfy the inequality
 \begin{equation}
     \xib^T \tilde{\mathsf{D}} \xib \geq \tilde {C} \| \xib \|^2
 \end{equation}
 for some constant $C>0$ with $\tilde{\mathsf{D}}(\xb,t) = (\sigma(\xb,t) + \epsilon \tilde{\sigma}(\xb,t))(\sigma(\xb,t) + \epsilon \tilde{\sigma}(\xb,t))^T.$ And an analogous statement holds for $\mathsf{D}$ alone with constant $C.$
\end{assumption}
The perturbed process
\begin{equation}
    \der \Xb^{\epsilon}_t = \bb(\Xb^{\epsilon}_t, t )\der t + \left[ \sigma(\Xb^{\epsilon}_t, t)  + \epsilon \tilde{\sigma}(\Xb^{\epsilon}_t, t) \right]  \der \Wb_{t},
    \label{eq:app_psde}
\end{equation}
has a corresponding Fokker-Planck equation
\begin{equation}
\label{eq:app_pfke}
\begin{aligned}
\partial_t \rho^\epsilon_t &= -\nabla \cdot [ (\bb(\cdot, t ) \rho_t^\epsilon ] + \frac12 \nabla \nabla : \left[ \tilde{\mathsf{D}}(\cdot,t) \rho_t^\epsilon \right],\\
&:= \mathcal{L}^{\epsilon} \rho_t^{\epsilon}
\end{aligned}
\end{equation}
with \begin{math}
\label{eq:pdiff_tensor_1}
\tilde{\mathsf{D}}(\xb,t) = (\sigma(\xb,t) + \epsilon \tilde{\sigma}(\xb,t))(\sigma(\xb,t) + \epsilon \tilde{\sigma}(\xb,t))^T.
\end{math}

Our goal is to establish a notion of equivalence between the aforementioned perturbed process and the score-shifted process. 
Recalling the definition $\Sigmasf(\xb,t) = \sigma(\xb,t) \Tilde{\sigma}^T(\xb,t) + \sigma^T(\xb,t) \Tilde{\sigma}(\xb,t)$, the score-shifted SDE is 
\begin{equation}
    \der \Xb^{s}_t = \left[\bb(\Xb^{s}_t, t ) - \frac{\epsilon}{2}\bigl( (\Sigmasf(\Xb^s_t,t) s(\Xb^s_t,t) + \nabla : \Sigmasf(\Xb^s_t,t)\bigr)\right]\der t +  \sigma(\Xb^{s}_t, t) \der \Wb_{t},
    \label{eq:app_ssde}
\end{equation}
where
\begin{math}
    \left[\nabla : \mathsf{G} \right]_i = \sum_{j} \frac{\partial}{\partial \xb_j} \mathsf{G}_{ij}
\end{math}
and 
\begin{math}
    \nabla \nabla : \mathsf{G}  = \sum_{ij} \frac{\partial^2}{\partial \xb_i \partial \xb_j} \mathsf{G}_{ij}.
\end{math}
The solution of the associated Fokker-Planck equation,
\begin{equation}
\begin{aligned}
\label{eq:app_sfke}
\partial_t \rho^s_t &= -\nabla \cdot [ (\bb(\cdot, t ) - \frac{\epsilon}{2}\bigl( (\Sigmasf(\cdot,t) s(\cdot,t) + \nabla : \Sigmasf(\cdot,t)\bigr) \rho_t^s ] + \frac12 \nabla \nabla : \left[ \mathsf{D}(\cdot,t) \rho_t^s \right] 
\end{aligned}
\end{equation}
with 
\begin{math}
\mathsf{D}(\xb,t) = \sigma(\xb,t)\sigma^T(\xb,t)
\end{math}
defines $s$, and the score is given by,
\begin{equation}
    \label{eq:app_instantaneous_score}
    s(\xb,t) = \nabla \log \rho^s_t(\xb).
\end{equation}
Using the fact that
\begin{equation}
    \Sigmasf \nabla \log \rho  + \nabla : \Sigmasf = \frac{\nabla : (\Sigmasf \rho)}{\rho},
\end{equation}
we deduce that
\begin{equation}
\partial_t \rho^s_t := \mathcal{L}^s \rho_t^s \equiv -\nabla \cdot [ (\bb(\cdot, t ) \rho_t^s ] + \frac12 \nabla \nabla : \left[ (\mathsf{D} + \epsilon \Sigmasf) \rho_t^s \right].
\end{equation}
Let $\chi:\Omega \to \RR$ be a test function. We are interested in the discrepancy
\begin{equation}
\begin{aligned}
&\int_{\Omega} \chi(\xb) \left( \rho_t^{\epsilon}(\der \xb) - \rho_t^s(\der \xb) \right), \\
= &\int_0^t \int_{\Omega} \chi(\xb) \left( \mathcal{L}^{\epsilon} \rho_{\tau}^{\epsilon}(\der \xb) - \mathcal{L}^s \rho_{\tau}^s(\der \xb) \right) d\tau, \\
= &\int_0^t \int_{\Omega} \chi(\xb) \left( \mathcal{L}^{s} (\rho_{\tau}^{\epsilon}(\der \xb) - \rho_{\tau}^s(\der \xb)) + \frac{\epsilon^2}{2} \nabla \nabla : \left[ \tilde{\sigma} \tilde{\sigma}^T \rho_{\tau}^\epsilon \right] \right) d\tau,\\
= &\int_0^t \int_{\Omega} \chi(\xb) \left( \left[ \mathcal{L}^{s} + \frac{\epsilon^2}{2} \tilde{\mathcal{L}} \right] \rho_{\tau}^{\epsilon}(\der \xb) - \mathcal{L}^s \rho_{\tau}^s(\der \xb)) \right) d\tau,\\
\end{aligned}
\end{equation}
where $\tilde{\mathcal{L}}$ is the diffusion operator with mobility $\tilde{\sigma}.$
The formal solution of the discrepancy can be written in terms of the time-order exponential operators associated with the linear PDE operators, i.e., 
\begin{equation}
    \delta \rho_t := \left(e^{t \left[\mathcal{L}^s + \tfrac{\epsilon^2}{2}\tilde{\mathcal{L}}\right]} - e^{t \mathcal{L}^s } \right) \rho_0 \equiv \rho_t^{\epsilon} - \rho_t^s.
\end{equation}
Letting $\epsilon \to 0$, we use the expansion,
\begin{equation}
    \delta \rho_t = \left( 1 + \frac{\epsilon^2}{2} t \tilde{\mathcal{L}} \right) e^{t \mathcal{L}^s} \rho_0 - e^{t \mathcal{L}^s } \rho_0 \implies \delta\rho_t = \frac{\epsilon^2}{2} t \tilde{\mathcal{L}} e^{t\mathcal{L}^s} \rho_0
    \label{eq:disc_exp}
\end{equation}
In particular~\cite{evans_partial_2010}, for each time $t>0$ there exists a constant $C_{\Omega}(t)$ depending only on $t, \Omega,$ and the coefficients of $\mathcal{L}^s$ such that 
\begin{equation}
    \mathrm{sup}_{\tau \in [0,t]} \| \rho_t^s \| \leq C_{\Omega}(t) \| \rho_0\|_{L_2(\Omega)}.
\end{equation}
In addition, the action of the linear operator $\tilde{\mathcal{L}}$ on the bounded density $\rho_t^s$ remains bounded by assumption. 
Hence, for any time $t$ we can choose $\epsilon > 0$ sufficiently small such that the scaled deviation
\begin{equation}
\epsilon^{-1} \int_{\Omega} \chi(\xb) \left( \rho_t^{\epsilon}(\der \xb) - \rho_t^s(\der \xb) \right)
\end{equation}
remains order $\epsilon$. 

\section{Fluctuation dissipation theorem for overdamped equilibrium systems subject to temperature changes}
\label{app:eq-od}

In this self-contained section, we derive the fluctuation dissipation relation for overdamped equilibrium dynamics subject to a change in the temperature, and prove the equivalence to the score sensitivity \eqref{eq:od3}. Without loss of generality, we consider the one-dimensional case
\begin{equation}
\label{eq:app-fdt1}
    \der X_t = -\nabla U(X_t)\der t + \sqrt{2 T}\der W_t
\end{equation}
with $X_0 \sim \rho^0(x) \propto \exp\left[-U(x)/T\right]$. At $t=0$, the temperature is perturbed to $T+\epsilon$. The instantaneous distribution $\rho_t$ of $X_t$ evolves according to the perturbed Fokker-Planck equation
\begin{equation}
\label{eq:app-fdt2}
    \partial_t \rho_t = \mathcal{L}\rho_t + \epsilon \mathcal{L}^1\rho_t
\end{equation}
with $\mathcal{L}f = -\nabla \left[-\nabla U f\right] + T \Delta f$ and $\mathcal{L}^1  = \Delta $. Following \cite{zwanzig_nonlinear_1973}, we expand $\rho_t = \rho^0_t + \epsilon \rho^1_t + \mathcal{O}(\epsilon^2)$ in powers of $\epsilon$ and solve \eqref{eq:app-fdt2} to each order:
\begin{equation}
\label{eq:app-fdt3}
    \begin{split}
        &\partial_t \rho^0_t = \mathcal{L}\rho^0_t\\
        &\partial_t \rho^1_t = \mathcal{L}\rho^1_t + \mathcal{L}\rho^0_t.
    \end{split}
\end{equation}
The formal first order solution of \eqref{eq:app-fdt2} is then given by
\begin{equation}
\label{eq:app-fdt4}
\rho_t = \rho^0 + \epsilon \int_0^t e^{\mathcal{L}(t-s)}\mathcal{L}^1 \rho^0 + \mathcal{O}(\epsilon^2),
\end{equation}
and sensitivities of bounded continuous observables read
\begin{equation}
\label{eq:app-fdt5_b}
    \left.\frac{\der \avg{A(t)}}{\der \epsilon}\right|_{\epsilon = 0} = \int_{\mathbbm{R}} \der x A(x)\int_0^t e^{\mathcal{L}(t-s)}\mathcal{L}^1 \rho^0. 
\end{equation}
Next, it is clear that $\mathcal{L}^1\rho^0 = - \beta^2(LU\rho^0)$, where $Lf = -\nabla U \nabla f+ T \Delta f $ is the adjoint of $\mathcal{L}$ and $\beta = T^{-1}$. In turn, denoting $\avg{\cdot}_0$ equilibrium expectations with respect to $\rho_0$, the sensitivity \eqref{eq:app-fdt5_b} is rewritten as

\begin{equation}
\label{eq:app-fdt5}
\begin{split}
    \left.\frac{\der \avg{A(t)}}{\der \epsilon}\right|_{\epsilon = 0} &= -\beta ^2\int_0^t \der s\int_{\mathbbm{R}} \der x L e^{L(t-s)}A(x) U(x)\rho^0\\
    &= \beta ^2\int_0^t \der s \frac{\der}{\der s} \avg{A(t)U(s)}_0\\
    &= \beta ^2 \left[\avg{\avg{A U}_0- A(t) U}_0\right]
    \end{split}
\end{equation}
and we recover the FDT \eqref{eq:od3}. We now show that one obtains the same result from the score sensitivity

\begin{equation}
\label{eq:app-fdt6}
    \partial_{\Tilde{\sigma}} A_t = \frac{1}{\sqrt{2 T^3}}\avg{A(t)\int_0^t\nabla U(X_s))\der W_s}
\end{equation}
with $\tilde{\sigma} = 1/\sqrt{2T}$. Using Îto calculus, we first compute the total differential of $U(X_t)$

\begin{equation}
\label{eq:app-fdt7}
    \der U(X_t) = \nabla U(X_t) \der X_t + \frac{1}{2} \Delta U(X_t) \der t = \nabla U(X_t) \left[-\nabla U(X_t) \der t + \sqrt{2 T}\der W_t\right]+ T \Delta U(X_t) \der t,
\end{equation}
such that

\begin{equation}
\label{eq:app-fdt8}
    \sqrt{2 T}\int_{0}^{t} \nabla U(X_s) \der W_s = U(X_t)-U(X_0) + \int_{0}^{t} \left[(\nabla U)^2-T\Delta U \right]\der s
\end{equation}
where we recognize once again the generator $L$ applied to the potential $U$

\begin{equation}
\label{eq:app-fdt9}
	\frac{1}{\sqrt{2 T^3}}\int_{0}^{t} \nabla U(X_s) \der W_s = \frac{1}{2T^2}\left[U(X_t)-U(X_0) - \int_{0}^{t} L U \der s\right].
\end{equation}
Using shorthand notations $B_s = B(X_s)$, we then obtain :

\begin{equation}
\label{eq:app-fdt0}
    \begin{split}
    \partial_{\Tilde{\sigma}} A_t &=\frac{1}{\sqrt{2 T^3}}\avg{A(t)\int_0^t\nabla U(X_s))\der W_s}\\
        &= \frac{1}{2 T^2}\left[\avg{A_tU_t}-\avg{A_tU_0} - \int_0^t \avg{A_t L U_s} \der s\right]\\
        &=\frac{1}{2 T^2}\left[\avg{A_tU_t}-\avg{A_tU_0} + \int_0^t \frac{\der}{\der s}\avg{A_t U_s} \der s\right]\\
        &=\frac{1}{2 T^2}\left[\avg{A_tU_t}-\avg{A_tU_0} + \left[\avg{A_tU_t}-\avg{A_tU_0}\right]\right]\\
        &=\frac{1}{T^2}\left[\avg{AU}-\avg{A_tU_0}\right]
    \end{split}
\end{equation}
and  recover the FDT \eqref{eq:app-fdt5}. For perturbed underdamped equilibrium dynamics

\begin{equation}
\label{eq:app-ud1}
     \begin{split}
        &\der \Xb_t = \Vb_t \der t,\\
         &\der \Vb_t = \left[- \nabla U(\Xb_t)-\gamma \Vb_t\right]\der t + \sqrt{2 \gamma (T+\epsilon)} \der \Wb_t, 
     \end{split}
 \end{equation}
where $\Xb_t$ and $\Vb_t$ describe the system's instantaneous positions and momenta, only the momenta part of the score $\nabla_{\vb}\rho_{\rm eq}(\xb, \vb) = - \vb /T$ is relevant to compute temperature sensitivities, and, in that case, the general formula \eqref{eq:sensfinal} reads

\begin{equation}
\label{eq:app-ud2}
\partial_{\tilde{\sigma}} \avg{A_t} = \frac{\gamma^{\frac{1}{2}}}{\sqrt{2 T^3}}\avg{A(t) \int_{0}^{t}\Vb_s \der \Wb_s},
\end{equation}
with $\tilde{\sigma} = 1/\sqrt{2\gamma T}$.

\section{Interacting active particles}

In this section, we provide details on the numerical implementation of the non equilibrium steady state system \eqref{eq:ness3}. In particular, we discuss the neural network (NN) representation and the numerical estimate of the stationary distribution.  We also gather here the values of parameters appearing in the dynamics \eqref{eq:ness3} and used throughout : $L = 10,\mu = 10,v_0 = 5, \gamma = 0.1,\alpha = 7.5,r = 1.0$. Finally, note that in the sensitivity framework~\eqref{eq:sensfinal}, static sensitivities to temperature variations for arbitrary observable $A$ correspond to $\partial_{\Tilde{\sigma}}A$ with

\begin{equation}
\tilde{\sigma}=
\left(\begin{smallmatrix}
    1/\sqrt{T} & 0 \\
    0 & 0 \\
\end{smallmatrix}\right).
\end{equation}

\subsection{Training details}
\label{app:active-training}
Since the NESS score $s^0$ is not analytically tractable, we use a neural network (NN) representation $\hat{s}$ trained on stationary simulation data, and minimize the agnostic/Fokker-Planck composite objective~\cite{Boffi2023}
\begin{equation}
\label{eq:app-loss2active}
\mathcal{L}\left[\hat{s}\right] = \mathbbm{E}_{\rho_{\rm ss}}\left[|\hat{s}|^2 + 2\nabla\hat{s}\right] + \mathbbm{E}_{\rho_{\rm ss}}\left[\left(\nabla.\hat{v} + \hat{v}.\hat{s}\right)^2\right].
\end{equation}

\paragraph{Loss function.}

We first comment on the composite loss \eqref{eq:app-loss2active} used as an objective for the score $\hat{s}$.
While the first term of the right hand side is the agnostic Hyvärinen \cite{hyvarinen_estimation_2005} score loss, the second term has first been introduced in \cite{Boffi2023} to take into account the dynamics of the system. Specifically, for Fokker Planck dynamics

\begin{equation}
\label{eq:app-nesstrain1}
    \partial_t \rho_t  = -\nabla\left[\bb \rho\right] + \frac12 \nabla \nabla : \left[ \mathsf{D}\rho_t \right]
\end{equation}
with space and time independent $\mathsf{D}$ and time independent $\bb$, the stationary score $s^0$ is shown to satisfy the following ODE:

\begin{equation}
\label{eq:app-nesstrain2}
    \nabla. v + v.s^0 = 0,\hspace{10pt}v =\bb - \frac{\mathsf{D}}{2} s^0.
\end{equation}
In turn, the objective \eqref{eq:app-loss2active} penalizes deviations from zero of the residual $\nabla \hat{v} + \hat{v}.\hat{s}$ , where $\hat{v} = \bb - \frac{\mathsf{D}}{2} \hat{s}$.  

\paragraph{Training protocol.} For each temperature $T$ in Fig. \ref{fig:ness}, we represent $\hat{s} : \mathbbm{R}^2 \to \mathbbm{R}^2$ as a fully connected NN with 4 layers, 256 hidden nodes and GELU activation function. The objective \eqref{eq:app-loss2active} is minimized on a training set of $10^5$ equilibrated samples obtained from simulation of the dynamics \eqref{eq:ness3}. We use the Adam \cite{kingma_adam_2015} optimizer with learning rate $\ell = 10^{-5}$, batch size 1024 and 100 learning epochs ($10^5$ learning steps).

\paragraph{Sensitivity estimation} The static score sensitivity $\partial_{\tilde\sigma}\avg{x^2}$ is averaged over $10^6$ trajectories of duration $t=4$ and time step $\der t = 10^{-3}$.

\subsection{Numerical solution to the Fokker Planck equation}
\label{app:active-nsolve}

Because of the low dimensionality of the SDE \eqref{eq:ness3}, we numerically solve the associated Fokker Planck equation

\begin{equation}
\label{eq:app-ness-num1}
\partial_t \rho_t(x,g) = -\partial_x \left[(2 \mu f(x) + v_0 g)\rho_t(x,g)\right]-\partial_g \left[(-\gamma g)\rho_t(x,g)\right] + \left[2 T \partial_{xx} + 2 \gamma \partial_{vv}\right]\rho_t(x,g)
\end{equation}
to obtain an alternative score estimate. Note that we solve \eqref{eq:app-ness-num1} on the finite domain $\left[-L / 2, L/2\right]^2$ and impose periodic boundary conditions; this approximation provides satisfactory results since the true stationary marginal of $g$ is a Gaussian with mean zero and variance 1. Because we know from simulation that the stationary distribution is bimodal (see figure \ref{fig:ness}), we initialize $\rho_{t=0}$ as a mixture of normal Gaussians centered at $(-L/4,0)$ and $(L/4,0)$ respectively. We use a pseudospectral integrator \cite{ceniceros_numerical_2004} on a grid of size $(512\times 512)$ for $4.10^4$ steps with time-step $\der t = 10^{-4}$. The fast Fourier transforms and inverse transforms are carried out using the python package \texttt{numpy} \cite{numpy}. Finally, we interpolate the discrete solution using \texttt{scipy}'s \cite{scipy} RegularGridInterpolator, and evaluate the corresponding score.

\paragraph{Sensitivity estimation} The static score sensitivity $\partial_{\tilde\sigma}\avg{x^2}$ in Fig. \ref{fig:ness} is averaged over $4\times 10^6$ trajectories of duration $t=4$ and time step $\der t = 10^{-3}$.

\section{Malliavin Calculus for the Black-Scholes model}
\label{app:malliavin}

In this section we show that the score sensitivity 
\begin{math}
    \avg{A(X_t)\left[\int_{0}^{t} \frac{W_s}{\sigma s}\der W_s- W_t\right]}
\end{math}
and Malliavin sensitivities \eqref{eq:bs-mav} \cite{Nualart, Fournie1999}
\begin{math}
    \avg{A(X_t)\left[\frac{W_t^2}{\sigma t} - W_t -\frac{1}{\sigma}\right]}
\end{math}
associated to variations of the volatility $\sigma$ in the Black-Scholes model are equal for all bounded $\mathcal{C}^1$ observables $A : \mathbbm{R} \to \mathbbm{R}$ and $t>0$, where the average is taken over the non stationary dynamics

\begin{equation}
\label{eq:app-bs2}
    \der X_t = \mu X_t \der t + \sigma X_t \der W_t;\hspace{10pt} X_0 = 1.
\end{equation}

\paragraph{Malliavin duality formula.} 
A thorough discussion on Malliavin calculus if well beyond the scope of this paper. However, we still need to introduce a few notations in order to prove that 
\begin{equation}
\label{eq:app-bs3}
	\avg{A(X_t)\left[\frac{W_t^2}{\sigma t} - W_t -\frac{1}{\sigma}\right]} = \avg{A(X_t)\left[\int_{0}^{t} \frac{W_s}{\sigma s}\der W_s- W_t\right]}.
\end{equation}
Fist, denoting $(\mathcal{F}_t)_{t\geq 0}$ the natural Brownian filtration  associated to $W_t$,  the Malliavin duality formula relates product expectations $\avg{A_t \int_0^t v_s \der W_s}$  of $(\mathcal{F}_t)_{t\geq 0}$ measurable processes $A_t$ and $v_t$  to the ``Malliavin derivative'' $D_s\left[A(X_t)\right]$ in the following sense

\begin{equation}
\label{eq:app-bsdual}
    \avg{A(X_t) \int_0^t v_s \der W_s } = \avg{\int_0^t \der s D_s\left[A(X_t)\right]v_s}.
\end{equation}
The Malliavin derivative is itself a complicated stochastic process which often does not have an explicit form.  However, In the Black-Scholes case \eqref{eq:app-bs2}, it is a.s. equal to

\begin{equation}
\label{eq:app-bs4}
D_s\left[A(X_t)\right] = \sigma A'(X_t)X_t
\end{equation}
for all $s\leq t$. We now make use of the duality formula \eqref{eq:app-bsdual} to rewrite the score sensitivity in terms of $A'(X_t)$. First

\paragraph{}
\begin{equation}
\label{eq:app-bs4-1}
	\begin{split}
	\avg{A(X_t)\left[\int_{0}^{t} \frac{W_s}{\sigma s}\der W_s- W_t\right]} 
	&=\avg{A(X_t)\left[\int_{0}^{t} \left(\frac{W_s}{\sigma s}-1\right)\der W_s\right]}\\
    &=\avg{A(X_t)\left[\int_{0}^{t}\der W_u \left(\int_{u}^{t}\frac{\der W_s}{\sigma s}-1\right)\right]}.
\end{split}
\end{equation}
We identify $v_u = \left(\int_{u}^{t}\frac{\der W_s}{\sigma s}-1\right)$ and use the Malliavin duality formula along with \eqref{eq:app-bs4} to obtain

\begin{equation}
\label{eq:app-bs5}
    \begin{split}
	\avg{A(X_t)\left[\int_{0}^{t} \frac{W_s}{\sigma s}\der W_s- W_t\right]} 
	&=\avg{A'(X_t)\sigma X_t \int_{0}^{t}\der u  \left(\int_{u}^{t}\frac{\der W_s}{\sigma s}-1\right)}\\
 &=\avg{A'(X_t)\sigma X_t \left(\int_{0}^{t}\frac{\der W_s}{\sigma s}  \int_{0}^{s}\der u-t\right)}\\
	&=\avg{A'(X_t)X_t\left(W_t-\sigma t\right)}\\
	\end{split}
\end{equation}
Finally, we use the fact (see \cite{Nualart}, chapter 6 for instance) that

\begin{equation}
\label{eq:app-bs6}
    \avg{A'(X_t)X_t\left(W_t-\sigma t\right)} = \avg{A(X_t)\left[\frac{W_t^2}{\sigma t} - W_t -\frac{1}{\sigma}\right]} 
\end{equation}
to conclude. We emphasize that this calculation leverages the specific explicit forms of both the score and Malliavin derivative associated to the Black-Scholes model. A more general statement to relate score and Malliavin sensitivities requires further investigation, and is left for future work.

\paragraph{Analytical sensitivity in Fig. \ref{fig:bs}.} Since $X_t = \exp\left[(\mu - \frac{\sigma^2}{2})t + \sigma W_t\right]$, the instantaneous distribution  $\rho_t$ of $X_t$ is given by

\begin{equation}
\label{eq:app-bs7}
    	\rho_t(x) = \exp\left[-\frac{\left[\log\left(x\right) - (\mu-\frac{\sigma^2}{2})\right]^2}{2 t \sigma^2}\right]/(\sqrt{2 \pi \sigma^2 t}s),
\end{equation}
and its second moment is explicit 
\begin{math}
\label{eq:app-bs7-1}
\avg{X_t^2} = e^{t(2\mu + \sigma^2)}
\end{math}
such that
\begin{equation}
    \partial_{\Tilde{\sigma}}\avg{X_t^2} = 2 t \sigma e^{t(2\mu + \sigma^2)}.
\end{equation}

\paragraph{Integrated observables.}

\begin{figure}
    \centering
    \includegraphics[width = \figsize\textwidth]{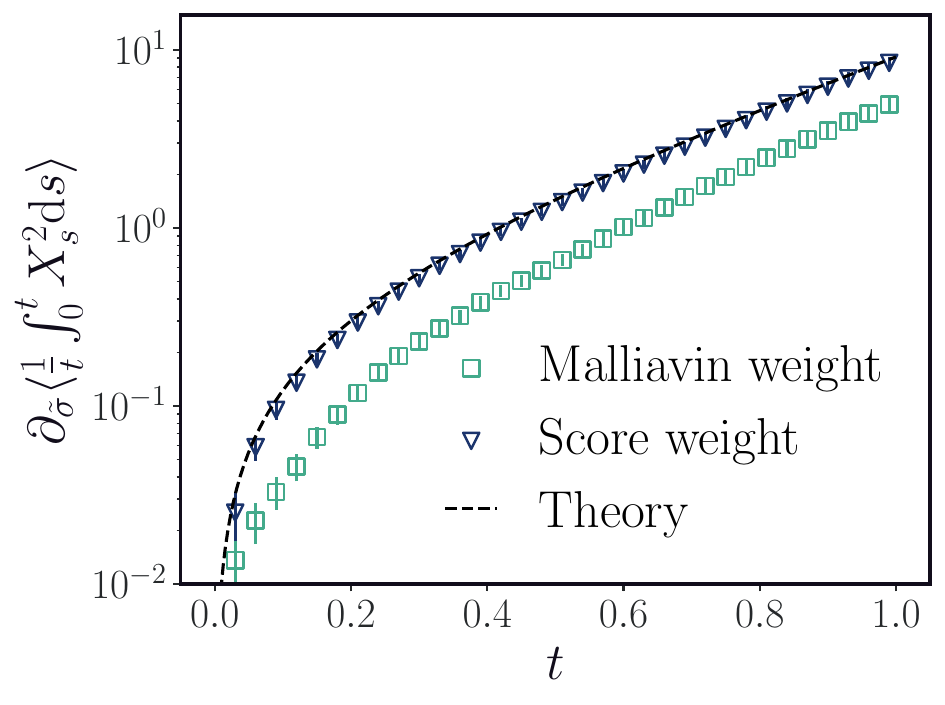}
    \caption{
    Black-Scholes model with $\mu=1,\sigma=1$; time dependent \textit{vega} of the time integrated observable $A_t = \left[\frac{1}{t}\int_0^t X_s^2 \der s \right]$. Both Malliavin and score estimates are averaged over $4.10^5$ sample trajectories. Whereas the score estimate is in perfect agreement with the dashed analytical prediction \eqref{eq:app-integrated-sens}, the Malliavin estimate \eqref{eq:app-bs6} does not hold for time integrated observables. }
    \label{fig:app-bs-discrepency}
\end{figure}

We emphasize that the score sensitivity formula holds for time integrated observables $A(\{X_t\}_{t\geq 0})$ which depend on the full trajectory, without the need for any modifications. In contrast, the Malliavin formalism doesn't carry through as smoothly. We showcase that discrepancy in the Black-Scholes model by comparing the analytical sensitivity of the time integrated observable $A_t = \left[\frac{1}{t}\int_0^t X_s^2 \der s \right]$ 

\begin{equation}
\label{eq:app-integrated-sens}
    \partial_{\tilde{\sigma}} A_t = \frac{2 \sigma   \left(e^{t \left(2 \mu +\sigma ^2\right)}
   \left(t \left(2 \mu +\sigma ^2\right)-1\right)+1\right)}{\left(2 \mu
   +\sigma ^2\right)^2}
\end{equation}
to the Malliavin estimate \eqref{eq:app-bs6} and  score estimate \eqref{eq:sensfinal} on Fig. \ref{fig:app-bs-discrepency}.

\bibliographystyle{unsrtnat} 
\bibliography{refs, references} 

\end{document}